\documentclass[nofootinbib,showpacs,preprintnumbers,amsmath,amssymb,superscriptaddress]{revtex4}

\usepackage{axodraw}
\usepackage{graphicx}

\begin{document}


\title{ Production of electroweak bosons
in  $e^+e^-$ annihilation at high energies }

\vspace*{0.3 cm}

\author{B.I.~Ermolaev}

\altaffiliation[Permanent Address: ]{Ioffe Physico-Technical Institute, 194021
 St.Petersburg, Russia}
\affiliation{CFTC, University of Lisbon
Av. Prof. Gama Pinto 2, P-1649-003 Lisbon, Portugal}
\author{S.M.~Oliveira}
\affiliation{CFTC, University of Lisbon
Av. Prof. Gama Pinto 2, P-1649-003 Lisbon, Portugal}
\author{S.I.~Troyan}
\affiliation{St.Petersburg Institute of Nuclear Physics, 188300 Gatchina, Russia}


\begin{abstract}
Production of electroweak bosons in  
$e^+e^-$ annihilation into quarks and into leptons 
at energies much greater than 100 Gev 
is considered. We account for double-logarithmic contributions to all
orders in electroweak couplings. 
It is assumed that the bosons are 
emitted in the multi-Regge kinematics. The explicit expressions  
for the scattering amplitudes of the process are obtained. It is shown that 
the cross sections of the photon and $Z$ production have the identical energy 
dependence and asymptotically 
their ratio depends only on the Weinberg angle 
whereas the energy dependence of 
the cross section of the $W$ production is suppressed 
by factor $s^{-0.4}$ compared to them.   
\end{abstract}


\maketitle

\section{Introduction}

Annihilation of $e^+e^-$ in the double-logarithmic approximation (DLA)
was considered first in
Ref.~\cite{ggfl}. In this work it was  shown that
when the total energy
of the annihilation is high enough, the most sizable radiative QED
corrections to $e^+e^-$ annihilation into $\mu^+ \mu^-$
are double-logarithmic (DL). These corrections were
calculated in Ref.~\cite{ggfl}  to all orders in $\alpha$. The DL
contributions to this process appear
when the final $\mu^+ \mu^-$ -pair is produced in the Regge kinematics,
i.e. when the muons move (in cmf) closely to the initial $e^+e^-$ -beam
direction.
According to the terminology introduced in Ref.~\cite{ggfl},
the process where $\mu^+$ moves in the $e^+ ~(e^-)$ -direction is
called  forward (backward) annihilation. 
Generalisation of these results to QCD (the forward and backward
annihilation of quarks into quarks of other flavours) 
and to the EW theory (the backward annihilation of the left handed leptons into
the right handed leptons) was obtained in Ref.~\cite{kl} and
Ref.~\cite{flmm} respectively. The forward and backward annihilation
of $e^+e^-$ into quarks, all chiralities accounted for, was
considered recently in Ref.~\cite{egt}. One of the
features obtained in Refs~\cite{ggfl}-\cite{egt}
is that the forward scattering
amplitudes in DLA are greater than the backward ones
in QED, in QCD and in EW theory.

Besides these $2 \to 2$, i.e.
elastic processes, it is interesting also to study the $ 2 \to 2 + n$
-exclusive processes accounting for emission of $n$ bosons
accompanying the elastic $2 \to 2$ annihilation.
The point is that besides the conventional, (soft) bremsstrahlung
there can be emitted harder
bosons. Emission of such bosons can be also studied in DLA to all orders
in the couplings, providing the hard bosons are emitted in cones
with opening
angles $\ll 1$ around the initial $e^+e^-$ beams, i.e. in the multi-Regge
kinematics. In this case, the most important part of the
inelastic scattering amplitudes
accounting for emission of $n$ bosons consists of the  kinematic factor
$ \sim (1/k_{1~\perp})\dots(1/k_{n~\perp})$ multiplied by
some function $M$ which is called the multi-Regge
amplitude of the process. The energy dependence of $M$ is controlled by
$n + 1$ electroweak Reggeons propagating in the crossing channel.
Description of the multi-Regge photon production in the backward
$e^+e^- \to \mu^+ \mu^-$ -annihilation was considered in Ref.~\cite{l}
and in Ref.~\cite{el1}. The multi-Regge amplitudes for gluon production 
 in the
backward annihilation
of quark-antiquark pairs were considered in
Ref.~\cite{el2}.

In the present paper we calculate
the scattering amplitudes for electroweak boson production in
$e^+e^-$ annihilation into quarks and leptons assuming that the
bosons are emitted in  the multi-Regge
kinematics. We use the approach of Refs.~\cite{el1},\cite{el2} and
account for electroweak double-logaritmic contributions to all orders
in the electroweak couplings. 
The paper is organised as follows: in Sect.~\ref{ONEBOSON} we
consider emission of one EW boson in $e^+e^-$ -annihilation into a
quark-antiquark pair.  We compose the infrared evolution equations
(IREE) for the amplitudes of these processes. The IREE  are solved in
Sect.~\ref{SOLUTION}~.  A generalisation of these results to the case
of emission of $n$ bosons is given in Sect.~\ref{NBOSONS}~. Emission of
the EW bosons in $e^+e$ annihilation into leptons is considered in
Sect.~\ref{toLEPTONS}~. Results of numerical calculations are presented
and discussed in Sect.~\ref{DISCUSS}~. Finally, Sect.~\ref{CONCLUSION}
is for conclusive remarks.

\section{Emission of one electroweak boson in the multi-Regge
kinematics}
\label{ONEBOSON}

Let us start by considering the process  
$e^+(p_2)e^-(p_1) \to q(p'_1)~ \bar{q}(p'_2)$  accompanied by emission of one  
electroweak boson with momentum $k$.  Energies of the bosons are 
assumed to be $ \gg M_Z$.  
There are two kinematics for this process that yield DL radiative
corrections. First of them is the kinematics where
$p'_1 \sim p_1$, $p'_2 \sim p_2$. Obviously,
\begin{equation}\label{tkin}
s = (p_1 + p_2)^2 \gg t_{1,2}~,\qquad t_1 = q^2_1 = (p'_1 - p_1)^2~,
\qquad t_2 = q^2_2 = (p_2 - p'_2)^2
\end{equation}
in this region. Eq.~(\ref{tkin}) means that the final
particles are in cones with opening angles $\theta \ll 1$ around the
$e^+e^-$ beams. 
The second kinematics is the one where
$p'_1 \sim p_2$, $p'_2 \sim p_1$ and therefore
\begin{equation}\label{ukin}
s = (p_1 + p_2)^2 \gg u_{1,2}~,\qquad u_1 = {q'}^2_1 =
(p'_2 - p_1)^2~,\qquad u_2 = {q'}^2_2 = (p_2 - p'_1)^2 .
\end{equation}

Eq.~(\ref{ukin}) means that the final
particles are also in cones with the cmf
opening angles $\pi- \theta \ll 1$ around the $e^+e^-$ beams.
Through this paper we call kinematics (\ref{tkin}) the $t$ -kinematics
and the kinematics (\ref{ukin}) - the $u$ -kinematics. Both of them
are of the Regge type
and studying them is similar in many
respects.

Instead of directly calculating inelastic amplitudes $A^{(\gamma, Z, W)}$
describing emission of any of $\gamma, Z, W$,
it is possible to calculate first the amplitudes
$A^{(0)}$ and $A^{(r)}$ ~$(r = 1,2,3)$ describing
emission of the isoscalar and the isovector bosons respectively. When
expressions for such amplitudes are obtained,
the standard relations between the fields $\gamma, Z, W$ and the fields
corresponding to the unbroken $SU(2)\otimes U(1)$ can be used in order
to express $A^{(\gamma, Z, W)}$ in terms of $A_0,~A_r$.  This way of
calculating $A^{(\gamma, Z, W)}$ is technically simpler than the direct
one because when the radiative corrections are taken into account in
DLA, contributions proportional to masses in propagators of all virtual
EW bosons are neglected and therefore both the isoscalar and the isovector
fields act as independent ones. It makes more convenient operating with
virtual isoscalar and isovector bosons than with $\gamma,Z,W$ -bosons.

It is also convenient to discuss a more general process
where lepton $l^i(p_1)$ (instead of $e^-$) and its antiparticle
$\bar{l}_{i'}(p_2)$ (instead of $e^+$)
annihilate into
the quark-antiquark pair $q^j(p'_1)~\bar{q}_{j'}(p'_2)$ and
a boson. The emitted boson can be either
the isoscalar boson $A_c$, with $c = 0$ or an isovector one
$A_c$,  with $c = 1,2,3$. We consider first the most difficult case when
 both $l^i$ and $q^j$ are left-handed particles, transitions to the other
chiralities are easy to do.
The scattering amplitude
of this process is
$ q_{j} \bar{q}^{j'} (M^c)^{i' j}_{i j'} \bar{l}_{i'} l^i$
where the matrix amplitude $(M^c)^{i' j}_{i j'}$ is the object to
calculate. In order to simplify operating with the isospin matrix structure
of $(M^c)^{i' j}_{i j'}$, it is convenient to regard the
process in the crossing channel, i.e.
in the $t$ -channel for kinematics (\ref{tkin}) and
in the $u$ -channel for kinematics (\ref{ukin}). When the process
$l^i \bar{l}_{i'} \to q^j \bar{q}_{j'} A_c$ is
considered in the $t$ -channel,
its amplitude can be expressed through the same matrix
$(M^c)^{i' j}_{i j'}$, however with initial (final) $t$-state
being  $q_{j} l^i (A_c \bar{q}^{j'} \bar{l}_{i'})$ :
\begin{equation}\label{mt}
M =  \frac{2}{k^2_{\perp}}\, A_c \bar{q}^{j'} \bar{l}_{i'}
(M^c)^{i' j}_{i j'}
 q_{j} l^i~.
\end{equation}

We have extracted the kinematic factor $2/k^2_{\perp}$ in order to
simplify the
matching condition (\ref{factor}) we will use.
The initial cross-channel
state  $q_j l^i$ in Eq.~(\ref{mt}) can be expanded into the sum
of the isoscalar and of the isovector irreducible $SU(2)$ representations:
\begin{equation}\label{initialt}
q_{j} l^i =
\left[\frac{1}{2}\, \delta^{i}_{j}\delta^{a}_{b} + 2\,
(t_m)^i_j(t_m)^b_{a} \right] q_b l^a~.
\end{equation}

The same is true for the final $\bar{q}^{j'}~\bar{l}_{i'}$ -pair. Therefore,
$(M^c)_{i j'}^{i' j}$ can be represented as the sum
\begin{equation}\label{mtinv}
(M^c)_{i j'}^{i' j} = \sum_{k = 0}^{4}(P^c_k)_{i j'}^{i' j} M_k ~.
\end{equation}
of the invariant amplitudes $M_k ~(k = 0,1,2,3,4)$, each multiplied the
projection operator $(P^c_k)_{i j'}^{i' j}$
corresponding to an irreducible $SU(2)$ -representation.
$k = 0,1$ correspond to emission of the isoscalar field and
$k = 2,3,4$ correspond to the isovector fields emission.
Then, the projection operator $(P^c_0)_{i j'}^{i' j}$ describes
the case (see Fig.~1) when both the initial $t$ -channel
fermion state and the final one
are $SU(2)$ singlets.

\begin{figure}[htbp]
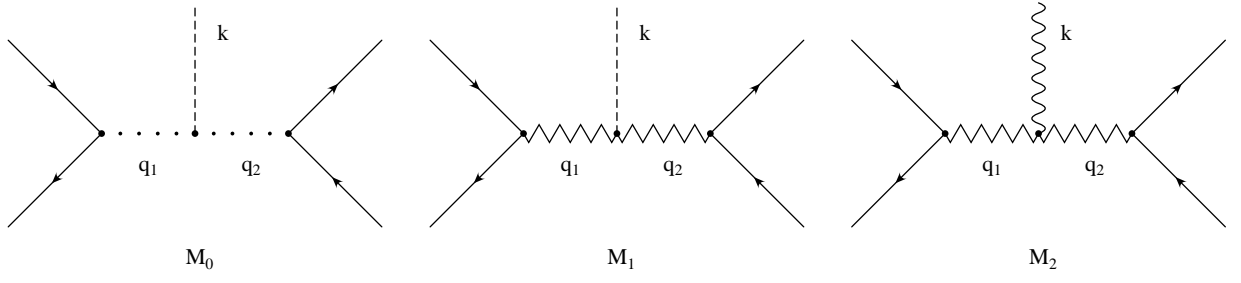

  \begin{center}
    \includegraphics[width=5cm]{images/Figura1.epsi}\hspace{.5cm}
    \includegraphics[width=5cm]{images/Figura1b.epsi}\hspace{.5cm}
    \includegraphics[width=5cm]{images/Figura1c.epsi}
   \caption{The multi-Regge invariant amplitudes $M_r$ (and the projector operators) in
    kinematics (1). The dotted lines correspond to the isoscalar
    Reggeons; whereas the zigzag lines stand for the isovector
    ones. The dashed lines denote isoscalar vector bosons and the
    waved line correspond to the isovector boson.}
    \label{fig.T1}
  \end{center}
\end{figure}

Obviously, in this case the emitted boson can be isoscalar only, i.e.
$c = 0$. Therefore
\begin{equation}\label{p0}
(P^c_0)^{i' j}_{i j'} = \frac{1}{2}
 \delta^c_0 \delta^{i'}_{j'} \delta^{j}_ {i} ~.
\end{equation}

The projection operators $P_k$, with $(k = 1,2)$
describe the cases when both the initial and the final $t$ -channel
states are the isovector $SU(2)$ states. However, $P_1$ corresponds to
the case when the emitted boson is isoscalar,
\begin{equation}\label{p1}
(P^c_1)^{i' j}_{i j'} =
2\, \delta^c_0 (t_m)^{j}_ {i}(t_m)^{i'}_{j'}
\end{equation}
whereas $P_2$ describes emission of isovector fields:
\begin{equation}\label{p2}
(P^c_2)^{i' j}_{i j'} =
(t_b)^{j}_ {i} (T^c)_{b a} (t_a)^{i'}_{j'} ~.
\end{equation}
$T^c$ (c = 1,2,3) in  Eq.~(\ref{p2}) stands for $SU(2)$ generators in
the adjoint (vector) representation.

Projector $P_3$ correspond to the case when the initial fermion
state is the $SU(2)$ singlet whereas the final one is the $SU(2)$ vector.
Projector $P_4$  describes the opposite situation. The
emitted boson is isovector in both these cases.
Therefore,
\begin{equation}\label{p34}
(P^c_3)^{i' j}_{i j'} =
 (t_c)^{j}_{i} \delta^{i'}_{j'}~,\qquad
(P^c_4)^{i' j}_{i j'} = \delta^{j}_{i} (t_c)^{i'}_{j'} ~.
\end{equation}

All operators  in Eqs.~(\ref{p0}, \ref{p34}) are orthogonal:
\begin{equation}\label{port}
(P_A)^{i' j}_{i j'} (P_B)_{i j'}^{i' j} \sim \delta^{AB} ~.
\end{equation}

Below (see Eq.~(\ref{factor})) we will show that
the invariant amplitudes $M_3, M_4$ do not have DL contributions. It
leaves us with amplitudes $M_{0,1,2}$ to calculate.
These invariant amplitudes account for radiative corrections to
all powers in the EW couplings in the DLA.
The arguments of $M_k$ are
\begin{eqnarray}\label{st}
s_1 &=& (p'_1 + k)^2 \approx 2p_1k~,\qquad
t_1 = q^2_1 = (p_1 - p'_1)^2,  \nonumber \\
s_2 &=& (p'_2 + k)^2 \approx 2p_2 k~~,\qquad
t_2 = q^2_2 = (p'_2 - p_2)^2~,
\end{eqnarray}
so that
\begin{equation}\label{s}
s_1\, s_2 = s\, k^2_{ \perp}
\end{equation}

The kinematics is the $t$ -channel multi-Regge kinematics when
\begin{equation}\label{s12}
s_{1,2} \gg t_{1,2} \geq M^2_Z~.
\end{equation}

Similarly, in order to simplify studying the isotopic structure of
$(M^c)_{i j'}^{i' j}$ of Eq.~(\ref{mt}) in kinematics
(\ref{ukin}), it is convenient to consider it in the $u$ -channel where
it can be expressed through $u$ -channel invariant amplitudes $M'_k$:
\begin{equation}\label{muinv}
(M^c)_{i j'}^{i' j} = \sum_{k = 0}^{2} ({P'}^c_k)_{i j'}^{i' j}\, M'_k ~.
\end{equation}

Operators $(P^{'c}_k)_{i j'}{i' j}$ describe irreducible $SU(2)$
representations, which for this channel are either symmetrical or
antisymmetrical two-quark states:
\begin{eqnarray}\label{p02u}
P^{'c}_0 &=& \frac{1}{2} \delta^c_0\left[
\delta^{i'}_{i}\delta^j_{j'} - \delta_{i}^{j} \delta^{i'}_{j'}\right] ~,\qquad
P^{'c}_1 = \frac{1}{2} \delta^c_0\left[
\delta^{i'}_{i}\delta^j_{j'} + \delta_{i}^{j} \delta^{i'}_{j'}\right] ~,
\nonumber \\
P^{'c}_2 &=& \frac{1}{2}\left[
\delta_i^{i'}(t^c)^j_{j'} + (t^c)_i^{i'}\delta^j_{j'} +
 \delta_{j'}^{i'} (t^c)_{i}^j + (t^c)_{j'}^{i'} \delta_{i}^j\right]~.
\end{eqnarray}

Operators $P'_0$, $P'_1$ describe emission of the isoscalar field
whereas $P'_2$ describes the isovector field emission.  Let us note
that  operator
\begin{equation}\label{p3'}
P^{'c}_3 =\frac{1}{2}\left[
\delta_i^{i'}(t^c)^j_{j'} + (t^c)_i^{i'}\delta^j_{j'}
 - \delta_{j'}^{i'} (t^c)_{i}^j - (t^c)_{j'}^{i'} \delta_{i}^j\right]
\end{equation}
must be accounted for\footnote{The gluon fields In Ref.~\cite{el2}
  were described in terms of matrixes $(t^cA^c)^a_b$ .}
 (cf Ref.~\cite{el2}).
However, the invariant amplitude $M'_3$ related to $P'$
does not yield DL contributions in the case of $SU(2)$ though it does 
in the case of $SU(N)$ with $N > 2$. It leaves us with three
invariant amplitudes $M'_{0,1,2}$  (just like it was in the case of the $t$
-kinematics). They depend on $s_{1,2}$ and on  $u_{1,2}$. In the
multi-Regge kinematics (\ref{ukin})
\begin{equation}
\label{s12u}
s_{1,2} \gg u_{1,2} \geq M^2_Z~.
\end{equation}

In order to specify the multi-Regge $t$ ($u$) -kinematics
completely, we assume that
\begin{equation}
\label{q12}
t_1 \gg t_2~,\qquad (u_1 \gg u_2) .
\end{equation}

The opposite case can be considered similarly. The kinematics where
$t_1 \sim t_2$ $(u_1 \sim u_2)$ means emission of soft electroweak
bosons. This kinematics will be considered below separately.

From the point of view of the Regge theory, accounting for radiative
corrections in kinematics (\ref{s12},\ref{s12u})
can be expressed through  exchange of Reggeons propagating in the
cross channels. Therefore operators $P_0$, $(P'_0)$ of
Eq.~(\ref{p0}) (Eq.~(\ref{p02u})) imply that
amplitude $M_0$ $(M'_0)$ is controlled by two isoscalar
Reggeons whereas the projection operators $P_{1,2}$ of
Eqs.~(\ref{p1},\ref{p2}) (operators $P'_{1,2}$ of  Eqs.~(\ref{p02u}))
imply that the energy dependence of amplitudes  $M_{1,2} (M'_{1,2})$ is
controlled by two isovector Reggeons.  In contrast to it, one of the
Reggeons in amplitudes $M_{3,4}$ is isoscalar and the other is
isovector.

Besides $s_{1,2}$ and $t_{1,2}$ $(u_{1,2})$, invariant
amplitudes $M_{0,1,2}$ $(M'_{0,1,2})$ depend also on the infrared (IR)
cut-off $\mu$ introduced in order to avoid IR singularities from
integrating over virtual particle momenta. We use the IR cut-off $\mu$
in the transverse space. However, definition of $\mu$ for radiative
amplitudes differs from the definition for elastic amplitudes.  In this
paper we introduce $\mu$ the same way as it was done in \cite{efl}. Let
us denote $k_{l~\perp}^{' a b}$ to be the component of a virtual particle
momenta $k'_l$ transverse to the plane
formed by momenta $a$ and $b$, with $a \neq b$.
Then, the IR cut-off $\mu$ obeys
\begin{equation}
\label{mu}
\mu < k_{l~\perp}^{' a b}
\end{equation}
for all $l = 1,\dots$ when $a,b = p_1, p_2, p'_1, p'_2, k$.
In the present paper we assume that $\mu \approx M_Z$.

In order to calculate $M_r$ we  generalise to the EW theory
the technique applied earlier to
investigation of the similar inelastic processes in QED\cite{el1} and in
QCD\cite{el2}. The essence of the method
is factorizing DL contributions from the virtual particles with minimal
$k_{l~\perp}^{' a b}$ and differentiating with respect to $\ln \mu^2$.
At $t_1, t_2 \gg \mu^2$, such particles can  only be
virtual EW bosons. Factorizing their contributions leads to the IREE for
amplitude $(M^c)_{i j'}^{i' j}$. This equation is depicted in Fig.~2.

\vspace{-10pt}\hfill \\
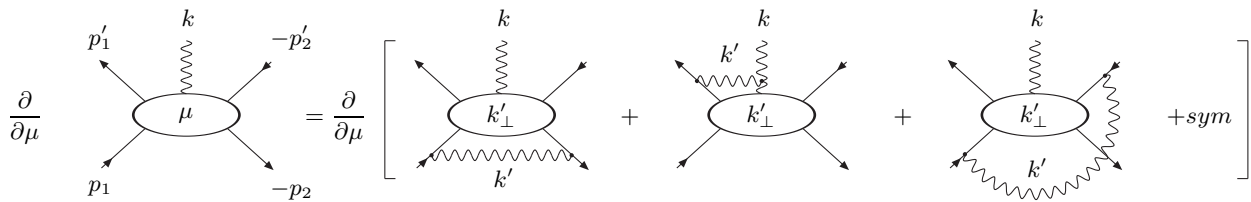
\begin{figure}[htbp]
\begin{center}
$\displaystyle{\frac{\partial}{\partial\mu}} $
\SetScale{0.8}
  \begin{picture}(90,60)(0,17)
    \Oval(60,25)(10,25)(0)
    \LongArrow(24,3.5)(24.1,3.6)
    \Line(20,0)(40.5,18.5)
    \LongArrow(40.5,31.5)(20,50)
    \LongArrow(97,47.35)(96.5,46.9)
    \Line(100,50)(79.5,31.5)
    \LongArrow(79.5,18.5)(100,0)
    \Photon(60,35)(60,60){2.5}{5}

    \Text(48,20)[]{$\mu$}
    \Text(20,-5)[rt]{$p_1$}
    \Text(80,-5)[lt]{$-p_2$}
    \Text(20,45)[rb]{$p'_1$}
    \Text(80,45)[lb]{$-p'_2$}
    \Text(49,60)[t]{$k$}
  \end{picture}
  $\displaystyle{=\frac{\partial}{\partial\mu}} $
  \begin{picture}(90,60)(0,17)
    \Line(5,-5)(5,60)
    \Line(5,-5)(10,-5)
    \Line(5,60)(10,60)
    \Oval(60,25)(10,25)(0)
    \LongArrow(24,3.5)(24.1,3.6)
    \Line(20,0)(40.5,18.5)
    \LongArrow(40.5,31.5)(20,50)
    \LongArrow(97,47.35)(96.5,46.9)
    \Line(100,50)(79.5,31.5)
    \LongArrow(79.5,18.5)(100,0)
    \Photon(60,35)(60,60){2.5}{5}
    \Photon(27,6)(93,6){-2.5}{10.5}
    \Vertex(27,6){1}
    \Vertex(93,6){1}

    \Text(48,20)[]{$k'_\perp$}
    \Text(49,60)[t]{$k$}
    \Text(49,0)[t]{$k'$}

 \end{picture}
$\displaystyle{+}$
  \begin{picture}(90,60)(05,17)
    \Oval(60,25)(10,25)(0)
    \LongArrow(24,3.5)(24.1,3.6)
    \Line(20,0)(40.5,18.5)
    \LongArrow(40.5,31.5)(20,50)
    \LongArrow(97,47.35)(96.5,46.9)
    \Line(100,50)(79.5,31.5)
    \LongArrow(79.5,18.5)(100,0)
    \Photon(60,35)(60,41){2.5}{1}
   \Photon(60,41)(60,60){-2.5}{3.5}
    \Photon(29.5,41)(60,41){2.5}{5}
    \Vertex(29.5,41){1}
    \Vertex(60,41){1}

    \Text(48,20)[]{$k'_\perp$}
    \Text(49,60)[t]{$k$}
    \Text(36.2,40)[b]{$k'$}
  \end{picture}
$\displaystyle{+}$
  \begin{picture}(90,60)(05,17)
    \Oval(60,25)(10,25)(0)
    \LongArrow(24,3.5)(24.1,3.6)
    \Line(20,0)(40.5,18.5)
    \LongArrow(40.5,31.5)(20,50)
    \LongArrow(97,47.35)(96.5,46.9)
    \Line(100,50)(79.5,31.5)
    \LongArrow(79.5,18.5)(100,0)
    \Photon(60,35)(60,60){2.5}{5}
    \PhotonArc(60,25)(37.396,210,30){2.5}{18.5}
    \Vertex(27,6.5){1}
    \Vertex(93,43.5){1}

    \Text(48,20)[]{$k'_\perp$}
    \Text(49,60)[t]{$k$}
    \Text(49,0)[]{$k'$}
  \end{picture}
$\displaystyle{+sym}$
  \begin{picture}(40,60)(05,17)
    \Line(10,-5)(10,60)
    \Line(10,-5)(5,-5)
    \Line(10,60)(5,60)
  \end{picture}
\end{center}
\vspace{10pt}
\caption{IREE for $M_Z$. Letters inside the blobs stand for infrared cutoffs.}
\end{figure}
\vspace{10pt}\hfill \\

Applying to it the projector operators of Eqs.~(\ref{p0} - \ref{p34},
\ref{p02u}) leads to the following IREE for $M_r$
(see Refs.~\cite{el1},\cite{el2}, \cite{egt} for technical details):
\begin{equation}
\label{eqm1}
\frac{\partial M_r}{\partial \rho_1} + \frac{\partial M_r}{\partial \rho_2} +
\frac{\partial M_r}{\partial y_1}
 + \frac{\partial M_r}{\partial y_2} =
-\frac{1}{8\pi^2} \big[ b_r \ln(s/ \mu^2) + h_r( y_1 + y_2) +
m_k y_1 \big] M_r ~.
\end{equation}

We have used in Eq.~(\ref{eqm1}) the fact that, according to
our assumption Eq.~(\ref{q12}), $k^2_{\perp} \approx t_1 (u_1)$
and introduced the logarithmic variables
\begin{equation}
\label{rho}
\rho_{1,2} = \ln(s_{1,2}/ \mu^2)~,\qquad y_{1,2} = \ln(t_{1,2}/ \mu^2)~.
\end{equation}

The numerical factors $b_r$, $h_r$ and $m_r$ in Eq.~(\ref{eqm1}) are:
\begin{eqnarray}
\label{bht}
b_0 &=& \frac{{g'}^2(Y - Y')^2}{4}~,\qquad
b_1 = b_2 = 2 g^2 +\frac{{g'}^2(Y - Y')^2}{4}~,\nonumber \\
h_0 &=& \frac{3g^2}{4} + \frac{{g'}^2 YY'}{4}~,\qquad
h_1 = h_2 = -\frac{g^2}{4} + \frac{{g'}^2 YY'}{4}~,\\
m_0 &=& m_1 = 0~,\qquad m_2 = g^2~.\nonumber
\end{eqnarray}

The IREE for the invariant amplitudes $M'_{0,1,2}$ can be obtained
similarly. It has the same structure as
Eq.~(\ref{eqm1}), though everywhere  $t_{1,2}$  should be
replaced by $u_{1,2}$. It means in particular that
$y_{1,2}$ should be replaced by $ y'_{1,2}= \ln (u_{1,2})/ \mu^2$.
This replacement and replacement of
operators $P_{0,1,2}$ by  $P'_{0,1,2}$ results in replacement of the
factors $b_r, h_r$ by $b'_r, h'_r$~:
\begin{eqnarray}
\label{bhu}
b'_0 &=& \frac{{g'}^2(Y + Y')^2}{4}~,\qquad
b'_1 = b'_2 = 2 g^2 +\frac{{g'}^2(Y + Y')^2}{4}~, \nonumber \\
h'_0 &=& \frac{3g^2}{4} - \frac{{g'}^2 YY'}{4}~,\qquad
h'_1 = h'_2 = -\frac{g^2}{4} - \frac{{g'}^2 YY'}{4}~.
\end{eqnarray}

Factors $m'_r$ in the IREE for $M'_r$
coincide with factors $m_r$ in Eq.~(\ref{bht}).
Therefore after replacements  $y_{1,2} \to y'_{1,2}$,
and $b_{0,1,2} \to b'_{0,1,2}$, $h_{0,1,2} \to h'_{0,1,2}$
Eq.~(\ref{eqm1}) for amplitudes $M_r$ in kinematics (\ref{s12})
holds for amplitudes $M'_r$ describing
$e^+e^- \to q\bar{q}$ -annihilation in kinematics (\ref{s12u}).

\section{Solving the evolution equations for $M_r$}\label{SOLUTION}

In order to solve Eq.~(\ref{eqm1}), it is convenient to operate with
the Mellin amplitude $F_r$
related to $M_r$ through the Mellin transform:
\begin{equation}
\label{f1}
M_r = \int_{-\imath \infty}^{\imath \infty} \frac{d \omega_1}{2 \imath \pi}
\frac{d \omega_2}{2 \imath \pi}
e^{\omega_1 \rho_1 + \omega_2 \rho_2} F_r(\omega_1, \omega_2, y_1, y_2) ~.
\end{equation}

In the $\omega$ -representation, multiplying by $\rho_i$ corresponds to
$-\partial / \partial \omega_i$.
Using this
and  Eqs.~(\ref{s},\ref{f1}), we can rewrite Eq.~(\ref{eqm1}) as
\begin{equation}
\label{eqf1}
\omega_1 F_r + \omega_2 F_r +
\frac{\partial F_r}{\partial y_1} + \frac{\partial F_r}{\partial y_2} =
b_r \Big( \frac{\partial F_r}{\partial \omega_1} +
\frac{\partial F_k}{\partial \omega_2} \Big)
+ \Big( \frac{1}{8 \pi^2}
\Big) \big[(b_r - h_r - m_r)y_1 - h_r y_2 \big] F_r~.
\end{equation}

For further simplification, it is convenient to introduce
variables $x_{1,2}$ and $z_{1,2}$~:
\begin{equation}
\label{xz}
x_{1,2} = \omega_{1,2}/ \lambda_r~,\qquad z_{1,2} = - \lambda_r y_{1,2}
\end{equation}
where $\lambda_r = \sqrt{b_r/ 8 \pi^2}$ ~.

In terms of $x_i, z_i$, the differential operator in the left hand side of
Eq.~(\ref{eqf1}) acquires symmetrical and simple form. Thus, we arrive at
\begin{equation}
\label{eqf1xz}
 \frac{\partial F_r}{\partial x_1} + \frac{\partial F_r}{\partial x_2} +
\frac{\partial F_r}{\partial z_1} + \frac{\partial F_r}{\partial z_2}
 = \big[(x_1 + x_2) + (1 + \beta_r)z_1  + \gamma_r z_2 \big] F_r ~
\end{equation}
where $\beta_r = - (h_r+m_r)/b_r$, $\gamma_r = - h_r/b_r$.

The general solution to Eq.~(\ref{eqf1xz}) can be written as
\begin{equation}
\label{f1gen}
F_r = \Phi_r (x_1 - z_2, x_2 - z_2, z_1 - z_2)\,
\exp\left[ \frac{x_1^2 + x_2^2}{2} +
(1 + \beta_r)\frac{z_1^2}{2} + \gamma_r \frac{z_2^2}{2}\right]
\end{equation}
where unknown function $\Phi_r$ has to be specified.
It can be done in particular through matching
\begin{equation}
\label{matchf1}
F_r (x_1,x_2,z_1,z_2)|_{z_2 = 0} = \tilde{F}_r(x_1,x_2,z_1)
\end{equation}
where $\tilde{F}_r$ is related through the Mellin transform
(\ref{f1}) to amplitude $\tilde{M}_r$
of the same process  in the
kinematics Eqs.~(\ref{s12},\ref{q12})
though with $q^2_2 \sim \mu^2$. The IREE (\ref{eqf1tilde})
for $\tilde{F}_r$ differs from the IREE of Eq.~(\ref{eqf1xz})
for $F_r$ in the following two respects. First, there is no $z_2$
dependence in Eq.~(\ref{eqf1tilde}). Second,  in contrast to
Eq.~(\ref{eqf1xz}), the IREE for $\tilde{F}_r$ contains an additional
term (that we denote $d Q_r(x)/d x$) (and will specify below in
Eq.~(\ref{f0})) in the rhs:
\begin{equation}
\label{eqf1tilde}
 \frac{\partial \tilde{F}_r}{\partial x_1} +
\frac{\partial \tilde{F}_r}{\partial x_2} +
\frac{\partial \tilde{F}_r}{\partial z_1}
 = \big[(x_1 + x_2) + (1 + \beta_r)z_1   +
\frac{d Q_r(x_2)}{d x_2} \big]\tilde{F}_r ~
\end{equation}

This new term corresponds to the situation when the particles with
the minimal transverse momenta are the $t_2$ -channel virtual quark pair
(see Fig.~3).
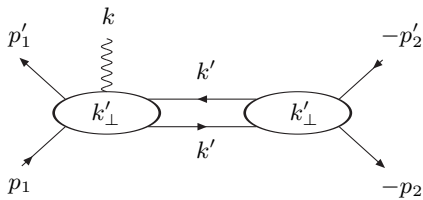
\begin{figure}[htbp]
\begin{center}
\SetScale{0.8}
  \begin{picture}(190,60)(0,0)
    \Oval(60,25)(10,25)(0)
    \LongArrow(24,3.5)(24.1,3.6)
    \Line(20,0)(40.5,18.5)
    \LongArrow(40.5,31.5)(20,50)
    \Photon(60,35)(60,60){2.5}{5}

    \Oval(150,25)(10,25)(0)
    \ArrowLine(130.5,31.5)(79.5,31.5)
    \ArrowLine(79.5,18.5)(130.5,18.5)
    \LongArrow(187,47.35)(186.5,46.9)
    \Line(190,50)(169.5,31.5)
    \LongArrow(169.5,18.5)(190,0)

    \Text(86,33)[b]{$k'$}
    \Text(86,10)[t]{$k'$}
    \Text(48,20)[]{$k'_\perp$}
    \Text(123,20)[]{$k'_\perp$}
    \Text(20,-5)[rt]{$p_1$}
    \Text(152,-5)[lt]{$-p_2$}
    \Text(20,45)[rb]{$p'_1$}
    \Text(152,45)[lb]{$-p'_2$}
    \Text(49,60)[t]{$k$}
  \end{picture}
\vspace{10pt}
   \caption{The soft fermion contribution to the IREE for
     $\tilde{M_Z}$}
  \end{center}
\end{figure}

This
contribution is $\mu$ -dependent only when $t_2 \approx \mu^2$.
The intermediate two-particle state in Fig.~3 factorizes
amplitude $\tilde{M}_r$ into a convolution of the same amplitude and the
elastic amplitude $E_r$.
The explicit expressions for the elastic electroweak amplitude
$E_r$ were obtained
in Ref.~\cite{egt}. The particular case where the produced
particles were a right handed lepton and its antiparticle was studied in
Ref.~\cite{flmm}. For all cases,
the Mellin amplitude $f_r$ is related to $E_r$
through the Mellin transform
\begin{equation}
\label{melline}
E_r = \int_{- \imath \infty}^{\imath \infty} \frac{d \omega}{2 \imath \pi}
e^{\omega \rho} f_r(\omega) ~,
\end{equation}
with $\rho = \ln(s/\mu^2)$,
can be expressed through the
Parabolic cylinder functions $D_{p_r}(x)$ with different values of $p_r$:
\begin{equation}
\label{f0}
f_r(x) = \frac{1}{p_r} \frac{d [\ln (e^{x^2/4}D_{p_r}(x))]}{dx} =
\frac{D_{p_r - 1}(x)}{D_{p_r}(x)}  \equiv ~\frac{1}{p_r}
\frac{d Q_r(x)}{d x}.
\end{equation}

The term $d Q_r(x_2)/d x_2$ in the rhs of  Eq.~(\ref{eqf1tilde})
corresponds to the contribution of the right blob in Fig.~3
to the IREE (\ref{eqf1tilde}).

The general solution to Eq.~(\ref{eqf1tilde})  is
\begin{equation}
\label{f1tildegen}
\tilde{F}_r = \tilde{\Phi}_r (x_1 - z_1, x_2 - z_1)
\frac{1}{D_{p_r}(x_2) \exp(x_2^2/4)}
\exp\left[ \frac{x_1^2 + x_2^2}{2} + (1 + \beta_r)\frac{z_1^2}{2}
\right]
\end{equation}
where there is, again, an unknown function $\tilde{\Phi}_r$.

In order to specify $\tilde{\Phi}_r$, we use the factorisation of
bremsstrahlung bosons with small
$k_{\perp}$ which takes place
(see Refs.~\cite{g},\cite{kl},\cite{efl},\cite{ce})
both in Abelian and in non-Abelian
field theories.
In the context of the problem under consideration it states that
when $z_1 = 0$, the radiative
amplitude $\tilde{M}_r$ and the
 elastic amplitude $E_r$ are related:
\begin{equation}
\label{factor}
\tilde{M}_r|_{z_1 = 0} = E_r G_r ~.
\end{equation}
$E_r$ in  Eq.~(\ref{factor}) are
the invariant amplitudes of the elastic annihilation process (see \cite{egt});
 $G_r = g'(Y \pm Y')/2$ for $r = 0$ (for invariant amplitude
$M_0 (M'_0)$ the sign is ``+'' (``-''));  $G_r = g$ for $r = 1,2,3$.

Eq.~(\ref{factor}) means that when $z_1 = 0$, the two Reggeons
in every amplitude $M_r$
converge into one
Reggeon that controls $E_r$ energy dependence.
However, such convergence is 
possible in the DLA only when both Reggeons are either isoscalar or 
isovector.
This rules amplitudes $M_{3,4}$ out of consideration. Obviously,
this property of the multi-Regge amplitudes
holds for the more complicated cases when the number of involved Reggeons is
more than two. This property was first obtained in Ref.~\cite{el2} and
was called  ``Reggeon diagonality''.

The matching (\ref{factor}) can be rewritten in terms of
Mellin amplitudes
$\phi_r(x_1, x_2) \equiv \tilde{F}_r(\omega_1, \omega_2, z_1)|_{z_1 = 0}$
and $f_r(\omega)$ :
\begin{equation}
\label{f1f0}
\int_{-\imath \infty}^{\imath \infty} \frac{d \omega}{2 \imath \pi}
e^{\omega_1 \rho_1 + \omega_2 \rho_2} \phi_r(\omega_1, \omega_2)
= G_r
\int_{- \imath \infty}^{\imath \infty} \frac{d \omega}{2 \imath \pi}
e^{\omega (\rho_1 + \rho_2)} f_r(\omega) ~.
\end{equation}

We have used in  Eq.~(\ref{f1f0})
that according to Eq.~(\ref{s}) $\rho = \rho_1 + \rho$  when $z_1 = 0$.
For the amplitudes with positive signatures that  we discuss
in the present paper,  the transform inverse to
Eq.~(\ref{f1}) can be written as
\begin{equation}
\label{inv}
M_r (\rho_{1,2}, y_{1,2}) = \int_0^{\infty} d \rho_1 d \rho_2
e^{- \omega_1 \rho_1 - \omega_2 \rho_2}
F_r(\omega_1, \omega_2, y_{1,2}) ~.
\end{equation}

Applying this transform to Eq.~(\ref{f1f0}) at $z_1 = 0$ leads to

\begin{eqnarray}
\label{phi1}
\phi_r(\omega_1, \omega_2) &=& G_r
\int_{- \imath \infty}^{\imath \infty} \frac{d \omega}{2 \imath \pi}
\int_0^{\infty} d \rho_1 d \rho_2
e^{(\omega - \omega_1) \rho_1 + (\omega - \omega_2) \rho_2}
f_r( \omega ) =   \nonumber \\
& & G_r
\int_{- \imath \infty}^{\imath \infty} \frac{d \omega}{2 \imath \pi}
\frac{1}{(\omega - \omega_1)(\omega - \omega_2)} f_r(\omega) ~.
\end{eqnarray}

Choosing the integration contours in Eq.~(\ref{phi1})
so that $\Re \omega < \Re \omega_{1,2}$ allows us to
integrate over $\omega$
by closing the contour to the right, which does not involve
dealing with singularities of $f_r$.
When this integration is done, we arrive at
\begin{equation}
\label{phi1f0}
\phi_r = \frac{f_r(\omega_1) - f_r(\omega_2)}{\omega_2 - \omega_1} ~.
\end{equation}

After that it is easy to obtain the following expression for $M_r$:
\begin{eqnarray}
\label{m1expr}
M_r &=& G_r \int_{-\imath \infty}^{\imath \infty}
\frac{d \omega_1}{2 \imath \pi}
\frac{d \omega_2}{2 \imath \pi}
\left(\frac{s_1}{q^2_1}\right)^{\omega_1}
\left(\frac{s_2}{\sqrt{q^2_1q^2_2}}\right)^{\omega_2}
\left[\frac{f_r(x_1) - f_r(x_2)}{\omega_2 - \omega_1}\right] \cdot
\nonumber \\
& & \frac{D_{p_r}(x_2 - z_1)}{D_{p_r}(x_2 - z_2)}
\exp \left[ - \frac{(1 - 2\beta_r)}{4}z^2_1
- \frac{(1 - 2 \gamma_r)}{4}z_2^2 \right] ~.
\end{eqnarray}

If we choose $\Re x_1 < \Re x_2$, Eq.~(\ref{m1expr}) takes simpler form:
\begin{equation}
\label{R}
M_r = G_r R_r
\end{equation}
where
\begin{eqnarray}
\label{f1order}
R_r &=&  \int_{-\imath \infty}^{\imath \infty}
\frac{d \omega_1}{2 \imath \pi}
\frac{d \omega_2}{2 \imath \pi}
\left(\frac{s_1}{q^2_1}\right)^{\omega_1}
\left(\frac{s_2}{\sqrt{q^2_1q^2_2}}\right)^{\omega_2}
\frac{1}{\omega_2 - \omega_1}\,
\frac{D_{p_r - 1}(x_1 - z_1)}{D_{p_r}(x_1 - z_1)} \cdot
\nonumber \\ & &
\frac{D_{p_r}(x_2 - z_1)}{D_{p_r}(x_2 - z_2)}
\exp \left[ - \frac{(1 - 2\beta_r)}{4}z^2_1
- \frac{(1 - 2 \gamma_r)}{4}z_2^2 \right] ~.
\end{eqnarray}

The amplitudes $M^{(\gamma)}, M^{(Z)}, M^{(W^{\pm})}$ of the electroweak
boson production are easily expressed through $R_r$:
\begin{eqnarray}
\label{mrt}
M^{(\gamma)} &=& \cos \theta_W M^{(0)} + \sin \theta_W M^{(3)} =
g\cos \theta_W (R_0 + R_1) + g \sin \theta_W R_2~, \nonumber \\
M^{(Z)} &=& -\sin \theta_W M^{(0)} + \cos \theta_W M^{(3)} =
-g\sin \theta_W (R_0 + R_1) + g\cos \theta_W R_2~, \nonumber \\
M^{(W^{\pm})} &=& (1/ \sqrt{2})[M^{(1)} \pm \imath M^{(2)}] =
(g/ \sqrt{2}) R_2
\end{eqnarray}
when the boson are produced in kinematics
(\ref{s12}). For kinematics (\ref{s12u}), the boson production
amplitudes are expressed through $R'_r$:
\begin{eqnarray}
\label{mru}
M^{(\gamma)} &=&
g\cos \theta_W (R'_0 + R'_1) + g \sin \theta_W R'_2~, \nonumber \\
M^{(Z)} &=& -g\sin \theta_W (R'_0 + R'_1) + g\cos \theta_W R'_2~,\nonumber \\
M^{(W^{\pm})} &=& (g/ \sqrt{2})R'_2~.
\end{eqnarray}

The exponent in  Eq.~(\ref{f1order}) is the Sudakov form factor for
this process. It accumulates the soft DL
contributions, with virtualities $\le z^2_1$. The harder DL contributions
are accounted through $D_p$ -functions. It is convenient to
perform integration over $\omega_{1,2}$ by taking residues. Such residues
are actually zeros $\bar{x}_k(r)~(k = 1,\dots)$
of involved $D_{p_r}$ -functions, so
\begin{equation}
\label{mrzeros}
R_r \sim \sum_{k = 1}
\left(\frac{s_1}{q^2_1}\right)^{\lambda_r \bar{x}_k(r)}
\left(\frac{s_2}{\sqrt{q^2_1q^2_2}}\right)^{\lambda_r \bar{x}_k(r)}~.
\end{equation}

Position of  $\bar{x}_k(r)$
depend on values of $p_r$ in such a way that the greater $p_r$, the
greater are $ \Re \bar{x}_k(r)$. In particular, the
real part of the rightmost zero $\equiv\bar{x}(r)$ is positive when
$p_r > 1$. In other words, $R_r$ increase with the
total energy when $p_r > 1$. Ref.~\cite{egt} states that
\begin{equation}
\label{pu}
p'_0 = \frac{3 - YY' \tan^2 \theta_W}{(Y + Y')^2 \tan^2 \theta_W},
\qquad~~p'_1 = p'_2 =
-\frac{1 + YY' \tan^2 \theta_W}{8 + (Y + Y')^2 \tan^2 \theta_W}
\end{equation}
and
\begin{equation}
\label{pd}
p_0 = \frac{3 + YY' \tan^2 \theta_W}{(Y - Y')^2 \tan^2 \theta_W},
\qquad~~p_1 = p_2 =
-\frac{1 - YY' \tan^2 \theta_W}{8 + (Y - Y')^2 \tan^2 \theta_W} .
\end{equation}

Therefore, only $M_0$ and $M'_0$ grow with increase of the annihilation 
energy whereas the 
amplitudes $M_{1,2}$ and ${M'}_{1,2}$ are falling.   

Let us discuss the asymptotics of $R_r$ first.
The asymptotics of the energy dependence of
each $R_r$ is controlled
by two identical isoscalar (isovector) leading Reggeons.
Intercepts $\Delta_j ~(j = S,V,S',V')$ of these Reggeons
are related to the position of the
rightmost zero $\bar{x}(j)$ of the $D_{p_r}$ -functions 
so that
\begin{equation}
\label{delta}
\Delta_S = \lambda_0\bar{x}(p_0)~,\qquad
\Delta_{S'} = \lambda'_0\bar{x}(p'_0)~,\qquad
\Delta_V = \lambda_1\bar{x}(p_1)~,\qquad
\Delta_{V'} = \lambda'_1\bar{x}(p'_1)~.
\end{equation}

We remind that $\lambda_j = \sqrt{b_j/8 \pi^2}$.
Therefore we arrive at the following asymptotics:
\begin{eqnarray}
\label{as}
M_0 &\sim& g' s^{\Delta_S}~,\qquad M_1 \sim g' s^{\Delta_V}~,\qquad
M_2 \sim g s^{\Delta_V}~,  \nonumber \\
M'_0 &\sim& g' s^{\Delta_{S'}}~,\qquad M'_1 \sim g' s^{\Delta_{V'}}~,\qquad
M'_2 \sim g s^{\Delta_{V'}}~.
\end{eqnarray}

As  the intercepts of the isoscalar Reggeons are greater than
the ones of the isovector Reggeons,
the asymptotics of
the exclusive cross sections $\sigma^{(\gamma)}$ and $\sigma^{(Z)}$
of the photon and $Z$ -production
is given by contributions of
the isoscalar Reggeons with intercepts  $\Delta_S$ and
$\Delta_{S'}$. Therefore,
the only difference between these
cross sections is the different couplings of these fields to the
isoscalar Reggeons. So, we conclude
(see Eqs.~(\ref{mrt}),(\ref{mru})) that asymptotically
\begin{equation}
\label{gammaz}
\frac{\sigma^{(Z)}}{\sigma^{(\gamma)}} \approx \tan^2 \theta_W ~.
\end{equation}

Accounting for contributions of other zeros, $\bar{x}^{(r)}$
changes the values of $\sigma^{(\gamma)}$ and $\sigma^{(Z)}$
but does not change Eq.~(\ref{gammaz}).
In contrast to Eq.~(\ref{gammaz}), the asymptotics of
the ratio $\sigma^{(\gamma)}/ \sigma^{(W)}$ depends on $s$.
The point is that the exclusive cross section $\sigma^{(W)}$
of $W$ -production involves the isovector Reggeons with smaller intercepts.
So, asymptotically this cross section
obeys the following relation:
\begin{equation}
\label{sigmawt}
\frac{\sigma^{(\gamma)}}{\sigma^{(W)}} \sim s^{2(\Delta_S - \Delta_V)}
 = s^{-0.36}~.
\end{equation}

However, contributions of other zeros of $D_{p}$ -functions change this
asymptotic relation. Results of numerical calculation of
$\sigma^{(W^{\pm})}$, $\sigma^{(Z)}$ and $\sigma^{(\gamma)}$,
and accounting for non-leading DLA contributions are discussed
in Sect.~\ref{DISCUSS}.

\section{Emission of $n$ vector bosons in the multi-Regge kinematics}
\label{NBOSONS}

The arguments of the previous Sects. can be extended in the 
straightforward way to the case when  
the $e^+e^-$ -annihilation into quarks is accompanied by 
emission of $n$ isoscalar or isovector bosons with momenta 
$k_1, ..., k_n$ 
in the multi-Regge kinematics. It is not 
difficult to generalise expressions of 
Eqs.~(\ref{p0}-\ref{p2}, \ref{p02u}) for projection operators to the 
case of the $n$ boson emission and obtain new projector operators.  
First of all, let us note that all non-diagonal projectors should be 
ruled out of consideration by the same reason as it was done in Sect.~1. 
Therefore, the invariant 
amplitudes of emission of $n$ bosons involve $n + 1$ identical 
intermediate Reggeons. The isotopic quantum numbers of the Reggeons 
depend on the initial fermion state and on the isospin of the emitted 
bosons.  
If the initial fermion state is isoscalar (or antisymmetric), the 
same is true for all intermediate Reggeons and therefore only 
isoscalar bosons can be emitted in these cases. The projector operators 
for this case are again 
$P_{0,1}$ 
($P'_{0,1}$) with 
trivial adding factors $\delta_{{c}_i 0}$ for every 
isoscalar boson. If the 
initial fermion state is isovector (or symmetrical), the emitted 
bosons can be both isoscalar or isovector gauge fields. Accounting for 
emission of the isoscalar bosons does not require any changes of the 
projectors. When  $r$ $(r \le n)$ 
isovector bosons $c_1, c_2,.., c_r$ are emitted in 
the $t$ -kinematics,  
the operators $P_2$ 
of Eq.~(\ref{p2}) should be replaced by 
\begin{equation}
\label{p'}
(P^{(c_1,c_2,\dots,c_n)}_2)^{i' j'}_{i j} =
(t_b)^{j'}_ {i'} (T^{c_r})_{b a_r} \dots
(T^{c_2})_{a_2 a_1}(T^{c_1})_{a_2 a_1} ({t_a}_1)^i_j ~.
\end{equation}

A similar generalisation of operator $P'_2$ is also easy to obtain.
The new invariant amplitudes $M_j(M'_j)$
corresponding to these operators depend on
 $n+1$ variables $s_i$:
\begin{equation}
\label{si}
s_1 = 2p_1 k_1~,\qquad s_2 = 2k_1 k_2~,\dots~,\qquad s_{n+1} = 2k_n p_2
\end{equation}
and on $t_i (u_i)$ in the case of the
t (u) - kinematics:
\begin{equation}
\label{tiui}
t_i = q^2_i~,\qquad u_i = {q'}^2_i
\end{equation}
where
\begin{eqnarray}
\label{qni}
q_1 &=& p'_1 - p_1~,\qquad q_2 = q_1 - k_1~,\dots~,\qquad
q_{n+1} = q_n - k_n = p_2 - p'_2  \nonumber \\
q'_1 &=& p'_2 - p_1~,\qquad q'_2 = q'_1 - k_1~,\dots~,\qquad
q'_{n+1} = q'_n - k_n = p_2 - p'_1~.
\end{eqnarray}

The kinematics is the multi-Regge $t$ -kinematics if
\begin{equation}
\label{sti}
s_i \gg t_j \geq M^2_Z
\end{equation}
and it is the multi-Regge $u$ -kinematics if
\begin{equation}
\label{sui}
s_i \gg u_j \ge M^2_Z ,
\end{equation}
with $i,j = 1,\dots~,n+1$. In order to define these kinematics completely, one
should fix relations between different $t_i$ (different $u_i$). In this
paper we consider the simplest
case of the monotonic ordering. We assume that
\begin{equation}
\label{orderti}
t_1 \gg t_2 \gg\dots\, \gg t_{n+1}
\end{equation}
for the case of the multi-Regge $t$ -kinematics
and the similar monotonic ordering
\begin{equation}
\label{orderui}
u_1 \gg u_2 \gg\dots\, \gg u_{n+1} ~
\end{equation}
for the case of the multi-Regge $u$ -kinematics
\footnote{Scattering amplitudes for other multi-Regge
kinematics can be calculated
similarly (see Ref.~\cite{el2}).  It is worth to mention that amplitudes
for kinematics
(\ref{orderti}) and (\ref{orderui}) yield main
contributions to the inclusive cross section when integration over the
EW boson momenta is performed.}.
Eqs.~(\ref{si},\ref{tiui}) read that in the both kinematics
\begin{equation}
\label{sn}
s_1\dots\, s_{n+1} = s k_{1~\perp}^2\dots\, k_{n ~\perp}^2~.
\end{equation}

It is also convenient to introduce variables $\rho_i = \ln(s_i / \mu^2)$ and
$y_i$ where $y_i = \ln (t_i/ \mu^2)$ for the forward kinematics
($y'_i = \ln (u_i/ \mu^2)$ for the case of the backward one).
In these terms, The IREE for $M^{(n)}_j$ looks quite similar to
Eq.~(\ref{eqm1}):
\begin{eqnarray}
\label{eqmn}
&&\frac{\partial M^{(n)}_j}{\partial \rho_1} + \dots\,+
\frac{\partial M^{(n)}_j}{\partial \rho_{n+1}} +
\frac{\partial M^{(n)}_j}{\partial y_1} + \dots\,
 + \frac{\partial M^{(n)}_j}{\partial y_{n+1}} =  \nonumber \\
&& - \frac{1}{8 \pi^2}
\left[ b_j  \ln(s/\mu^2) - h_j( y_1 + y_{n+1}) + \sum_l m_l y_l \right]
M^{(n)}_j ~,
\end{eqnarray}
where $b_j, h_j$ are given by Eqs.~(\ref{bht},\ref{bhu}); $m_l = g^2$ if
the boson $l$ with momentum $k_l$ is isovector, otherwise $m = 0$.
Let us consider for simplicity the case of emission of isoscalar bosons.
Introducing the Mellin amplitude $F_n$ through the transform
\begin{equation}
\label{fn}
M_n = \int_{-\imath \infty}^{\imath \infty}
\frac{d \omega_1..d\omega_{n+1}}{2 \pi \imath}
e^{\omega_1 \rho_1 +\dots+ \omega_{n+1} \rho_{n+1}}
F_n(\omega_1,\dots, \omega_{n+1}, y_1,\dots, y_{n+1})
\end{equation}
and using notations $x_i, z_i$ defined as
\begin{equation}
\label{xzi}
x_i = \omega_i / \lambda~,\qquad z_i = - \lambda y_i ~,
\end{equation}
we transform Eq.~(\ref{fn}) into the following one:
\begin{eqnarray}
\label{eqfn}
&& \frac{\partial F_n}{\partial x_1} + \dots+
\frac{\partial F_n}{\partial x_{n+1}} +
\frac{\partial F_n}{\partial z_1} + \dots
+ \frac{\partial F_n}{\partial z_{n+1}}
= \nonumber \\
&& \left[(x_1+\dots+x_{n+1}) + z_1 +\dots + z_n  + h(z_1 + z_{n+1}) \right]
F_n ~,
\end{eqnarray}
with the general solution
\begin{eqnarray}
\label{fngen}
F_n &=&
\Phi_n \left(x_1 - z_{n+1}, x_2 - z_{n+1},\dots, x_{n+1} - z_{n+1};
z_1 - z_{n+1},\dots,z_n - z_{n+1}\right) \cdot \nonumber \\
& & \exp\left[S_{n+1}(x) + S_n(z) + h (\frac{z_1^2 +
z_{n+1}^2}{2})\right]
\end{eqnarray}
where  we have denoted $S_r(a) \equiv \sum_1^r a^2_i/2$ ~.
An unknown function $\Phi_n$ can be specified through the matching
\begin{equation}
\label{matchfn}
F_n|_{z_{n+1} = 0} = \tilde{F}_n
\end{equation}
where the Mellin amplitude $\tilde{F}_n$ describes
the same process in the multi-Regge kinematics
(\ref{si},\ref{qni},\ref{orderti}) though with $q^2_{n+1} = \mu^2$.
The IREE for $\tilde{F}_n$ is
\begin{eqnarray}
\label{eqfntilde}
&& \frac{\partial \tilde{F}_n}{\partial x_1} +\dots+
\frac{\partial \tilde{F}_n}{\partial x_{n+1}} +
\frac{\partial \tilde{F}_n}{\partial z_1} +\dots+
\frac{\partial \tilde{F}_n}{\partial z_{n+1}} =  \nonumber \\
&& \left[(x_1 +\dots+ x_{n+1}) + z_1 + \dots + z_{n+1} + h z_1  +
\frac{ d Q_r(x_{n+1})}{d x_{n+1}} \right]\tilde{F}_1
\end{eqnarray}
where $Q$ is defined by Eq.~(\ref{f0}).
The general solution to  Eq.~(\ref{eqfntilde}) can be
obtained quite similar to the one of  Eq.~(\ref{eqfn}):
\begin{eqnarray}
\label{fntilde}
\tilde{F}_n
&=& \tilde{\Phi}_n \left(x_1 - z_n, x_2 - z_n,\dots,x_{n+1} - z_n;
z_1 - z_n,\dots,z_{n-1} - z_n\right) \cdot \nonumber \\
& & \exp[ S_{n+1}(x) + S_n(z) + h \frac{z_1^2}{2}] ~.
\end{eqnarray}

It also contains an unknown function $\tilde{\Phi}$.
In order to specify it we use factorisation
of the photons with small $k_{\perp}$:
\begin{equation}
\label{matchmn}
\tilde{M}_n|_{z_n = 0} =
\tilde{M}_{n-1}(s_1,\dots,s_n; q^2_1,\dots,q^2_{n-1}, \mu^2)~.
\end{equation}

Rewriting this equation in terms of the Mellin amplitudes and
performing the transform inverse
to   Eq.~(\ref{fn}), we express $\tilde{F}_n$ through amplitude
$\tilde{F}_{n-1}$:
\begin{eqnarray}
\label{fnminus}
\tilde{F}_n|_{z_n = 0} &=& \frac{1}{(\omega_n - \omega_{n+1})}
\left[\tilde{F}_{n-1}(x_1,\dots,x_{n-1},x_n; z_1,\dots,z_{n-1}) - \right.
\nonumber \\           & & \left.
\tilde{F}_{n-1}(x_1,\dots,x_{n-1},x_{n+1}; z_1,\dots,z_{n-1}) \right] ~.
\end{eqnarray}

Combining this equation with Eqs.~(\ref{fn},\ref{fntilde}) leads to the
following recurrent
formula for $F_n$:
\begin{eqnarray}
\label{rec}
&& F_n(x_1,\dots,x_{n+1}; z_1,\dots,z_{n+1}) =
\frac{1}{\omega_n - \omega_{n+1}} \tilde{F}_{n-1}
\left(x_1 - z_n,\dots,x_n -z_n; z_1 -z_n,\dots,z_{n-1} - z_n \right) \cdot
\nonumber \\
&& \frac{D_p(x_{n+1} - z_n)}{D_p(x_{n+1} - z_{n+1})}
\exp\left[{S_{n+1}(x) - S_{n+1}(x - z_n) + S_n(z) - S_{n-1}(z-z_n) +
\frac{h}{2}[z^2_1 + z^2_{n+1} - (z_1 - z_n)^2]} \right]~.
\end{eqnarray}

Using this formula leads to the following expression for the amplitude
$M^{(n)}_j$ of emission of $n$ isoscalar bosons in the ordered kinematics
(\ref{orderti},\ref{orderui}):
\begin{eqnarray}
\label{fnorder}
M_r^{(n)} &=& \Big(g' \frac{(Y \pm Y')}{2}\Big)^n 
\int_{-\imath \infty}^{\imath \infty}
\frac{d \omega_1}{2 \imath \pi} \dots
\frac{d \omega_{n+1}}{2 \imath \pi}
\left(\frac{s_1}{q^2_1}\right)^{\omega_1} \dots
\left(\frac{s_{n+1}}{\sqrt{q^2_nq^2_{n + 1}}}\right)^{\omega_{n + 1}}
\frac{D_{p_r - 1}(x_1 - z_1)}{D_{p_r}(x_1 - z_1)} \cdot \nonumber \\
& & \frac{D_{p_r}(x_2 - z_1)}{D_{p_r}(x_2 - z_2)} \dots
\frac{D_{p_r}(x_{n+1} - z_n)}{D_{p_r}(x_{n+1} - z_{n+1})}
\exp \left[ - \frac{(b_r - 2h_r)}{4b_r}(z^2_1 + z_2^2) \right] ~.
\end{eqnarray}
where $Y \pm Y'$ corresponds to the kinematics  
(\ref{orderti},\ref{orderui}) respectively. 
Eq.~(\ref{fnorder}) implies that the contours of integrations obey 
$ \Re x_1 <.. < \Re x_{n + 1}$. After that one can perform integration 
in Eq.~(\ref{fnorder}) by taking residues in the $D_{p_r}$ zeros. 

When $k$ of the isoscalar bosons are replaced by the isovector ones, 
$(g'(Y \pm Y')/2)^n$ in Eq.~(\ref{fnorder}) should be replaced 
by $g^k(g'(Y \pm Y')/2)^{n - k}$; the factor     
$(m/2b_r) z^2_l$ for each of the emitted isovector bosons 
should be added to the last exponent. 
Using the standard relation between gauge fields $A_r$ and 
$\gamma, Z, W$, one can easily rewrite the 
gauge boson production amplitudes of Eq.~(\ref{fnorder}) 
in terms of amplitudes for the electroweak bosons production. 
 Asymptotics of the 
scattering amplitudes of  
the photon and $Z$ -production are governed by  the isoscalar Reggeons 
whereas $W$ -production involves the isovector Reggeons.    
Eq.~(\ref{fnorder}) can be used for obtaining different 
relation between cross sections of different radiative processes in the 
multi-Regge kinematics.  For example, 
\begin{equation}
\label{gammazn}
\frac{\sigma^{(nZ)}}{\sigma^{(n\gamma)}} \approx \tan^{2n} \theta_W ~.
\end{equation}
whereas the energy dependence of the
ratio $\sigma^{(n \gamma)}/\sigma^{(n W)}$ is less trivial.
Asymptotically,
\begin{equation}
\label{gammawn}
\frac{\sigma^{(n \gamma)}}{\sigma^{(n W)}} \sim s^{-0.36}.
\end{equation}

Results of accounting for the non-leading Reggeon contributions for
$\sigma^{(n \gamma)}/\sigma^{(n W)}$ can be obtained from Fig.~6
because
$\sigma^{(n \gamma)}/\sigma^{(n W)} = (e\sqrt{2}/g)^{n - 1}
\sigma^{(\gamma)}/\sigma^{( W)} $.  In obtaining
Eqs.~(\ref{gammazn},\ref{gammawn}) from Eq.~(\ref{fnorder})
we have used that according to Eq.~(\ref{sn}),
$(s_1)^{\Delta}\dots s_{n+1}^{\Delta} \sim s^{\Delta}$ .

\section{Emission of EW bosons in $e^+e^-$ -annihilation into leptons}
\label{toLEPTONS}

Inelastic annihilation of  $e^+e^-$, with the $e^-$ being left handed, 
into another left handed 
lepton $l'$ (i.e. into $\mu$ or $\tau$) and its 
antiparticle $\bar{l'}$ can be considered quite 
similarly to the annihilation into 
quarks discussed above.
In particular, explicit expressions for the new invariant 
amplitudes, $L_r^{(n)}$ of this process
\footnote{the kinematic factor $2/k_{\perp}$ is also extracted from $A_r^{(n)}$
like it was done for amplitudes  $M_r^{(n)}$. }
 can be 
obtained from Eqs.~(\ref{m1expr}),(\ref{R}),(\ref{fnorder}) 
by putting $Y' = Y = -1$ in Eqs.~(\ref{bht}),(\ref{bhu}) for the factors 
$b_r, h_r, b'_r, h'_r$.  
However having done it, we obtain that $b_0 = 0$ (see Eq.~(\ref{bht})). 
It means that the IR 
evolution equations for the scattering amplitude $L_0^{(n)}$ 
of the inelastic annihilation 
$e^+e^- \to l' \bar{l'} + n_1\gamma + (n - n_1) Z $  in the 
kinematics (\ref{tkin}) do not contain 
contributions proportional to $\ln(s/ \mu^2)$ in the rhs and therefore the 
Mellin amplitudes $f_0^{(n)}$ 
(related to $L_0$ through the Mellin transform (\ref{f1}), Eq.~(\ref{eqa1})) 
do not have the 
partial derivatives with respect to $\omega_j$ (cf Eq.~(\ref{fnorder})). 
In order to obtain expressions for the 
new scattering amplitudes $L_0^{(n)}$, let us consider first
the simple case of emission of one isoscalar boson accompanying the forward  
$e^+e^- \to l' \bar{l'}$ -annihilation, assuming that both $e^-$ and 
$l'$ are left particles. It is obvious that for this case,    
the IREE of Eq.~(\ref{eqm1})   
for scattering amplitude $M_0$ has to be replaced by the 
simpler one,  
\begin{equation}
\label{eqa1}
\frac{\partial L^{(1)}_0}{\partial \rho_1} + 
\frac{\partial L^{(1)}_0}{\partial \rho_2} + 
\frac{\partial L^{(1)}_0}{\partial y_1}
 + \frac{\partial L^{(1)}_0}{\partial y_2} = 
-\frac{1}{8\pi^2} \bar{h}_0( y_1 + y_2) L^{(1)}_0 ~
\end{equation}
where we have denoted $\bar{h}_0 = (3g^2 + {g'}^2)/4$ .  

In terms of the Mellin amplitude $f^{(1)}_0$,  Eq.~(\ref{eqa1})  
takes the following form:
\begin{equation}
\label{eql1}
(\omega_1 + \omega_2) f^{(1)}_0 +
\frac{\partial f^{(1)}_0}{\partial y_1} + 
\frac{\partial f^{(1)}_0}{\partial y_2} =  
\bar{h}_0 (y_1 +  y_2) f^{(1)}_0~. 
\end{equation}

The solution to Eq.~(\ref{eql1}) respecting the matching condition 
(\ref{f1}) is (cf Eq.~(\ref{m1expr}):     
\begin{eqnarray}
\label{l1}
L^{(1)}_0 = \frac{g'Y}{2} \int_{-\imath \infty}^{\imath \infty} 
\frac{d \omega_1}{2 \imath \pi}
\frac{d \omega_2}{2 \imath \pi}  
\Big(\frac{s_1}{q^2_1}\Big)^{\omega_1}
\Big(\frac{s_2}{q^2_2}\Big)^{\omega_2}   
\frac{(\bar{f}_0(x_1) - \bar{f}_0(x_2))}{(\omega_2 - \omega_1)} \nonumber \\
e^{(1/8\pi^2) \bar{f}_0(\omega_2)(y_1 - y_2)}  
e^{-(\bar{h}_0/2)(y^2_1 + y^2_2)} ~    
\end{eqnarray}
where 
the Mellin amplitude $\bar{f}_0 = $ 
for the elastic $e^+e^- \to \mu^+ \mu^-$ annihilation is 
(see \cite{kl,flmm,egt}) 
\begin{equation}
\label{fleptel}
\bar{f}_0 = 4 \pi^2 \left[\omega - 
\sqrt{\omega^2 - 
(3g^2 + {g'}^2)/8\pi^2} \right] ~. 
\end{equation}

The last exponent in  Eq.~(\ref{l1}) is the Sudakov form factor accumulating 
the DL contribution of the soft virtual EW bosons only. 
The other terms in the integrand account for harder contributions. The leading 
singularity (intercept), $\omega_0$  
of the integrand of  Eq.~(\ref{l1}) is given by the position of the 
branch point of the 
rhs of   Eq.~(\ref{fleptel}). Therefore we obtain    
\begin{equation}
\label{omega0}
\omega_0 = \sqrt{\frac{\alpha}{2 \pi}
\Big(\frac{3}{\sin^2 \theta_W} + \frac{1}{\cos^2 \theta_W} \Big)} = 0.13 ~,  
\end{equation}
so asymptotically 
\begin{equation}
\label{asymptlept1}
L_0 \sim s^{0.13} ~. 
\end{equation}
                   
The invariant amplitudes $L_0^{(n)}$ for production of $n$ 
isoscalar bosons in the kinematics (\ref{orderti}) 
when $e^+e^-$ annihilate into another lepton pair can be obtained similarly. 
The IREE for the amplitudes $L_0^{(n)}$ is 
\begin{equation}
\label{eqln}
\Big[\frac{\partial L^{(n)}_0}{\partial \rho_1} + ...+ 
\frac{\partial L^{(n)}_0}{\partial \rho_{n+1}} \Big] + 
\Big[\frac{\partial L^{(n)}_0}{\partial y_1}
 + ...+ \frac{\partial L^{(n)}_0}{\partial y_{n+1}} \Big] = 
-\frac{1}{8\pi^2} \bar{h}_0( y_1 +...+ y_{n+1}) L^{(n)}_0 ~
\end{equation}
and its solution is 
\begin{eqnarray}
\label{ln}
L^{(n)}_0 = \Big(\frac{g' Y}{2} \Big)^n \int_{-\imath \infty}^{\imath \infty} 
\frac{d \omega_1}{2 \imath \pi} ...
\frac{d \omega_{n+1}}{2 \imath \pi}  
\Big(\frac{s_1}{q^2_1}\Big)^{\omega_1}
\Big(\frac{s_2}{q^2_1}\Big)^{\omega_2}....
\Big(\frac{s_{n+1}}{q^2_n}\Big)^{\omega_{n+1}}   
\frac{\bar{f}_0(\omega_1)}
{(\omega_{n+1} - \omega_n) ...(\omega_2 - \omega_1)} \nonumber \\
e^{(1/8\pi^2) \left[\bar{f}_0(\omega_2)(y_1 - y_2) + ... 
+  \bar{f}_0(\omega_{n+1})(y_n - y_{n+1})\right]}  
e^{-(\bar{h}_0/2)(y^2_1 + ..+ y^2_{n+1})} ~    
\end{eqnarray}
when $\Re(\omega_i) < \Re(\omega_{i+1}), i = 1,..,n$ .  
Their asymptotic $s$-dependence is also given by  Eq.~(\ref{asymptlept1}). 
The results of numerical calculations for the cross section of $\gamma, Z$ 
and $W$ production in $e^+e^- \to l' \bar{l'}$ are presented in Fig.~6. 

\section{Numerical results}
\label{DISCUSS}

In order to estimate at what energy scale one might hope to observe
the predicted asymptotical behaviour of cross sections of exclusive
$W^{\pm}$ and $Z, \gamma$ production we have first to account for all
non-leading DLA amplitudes for left and right chiralities of initial
$e^+e^-$ and final $q\bar{q}$ or $l\bar{l}$ pairs. There are many such
amplitudes, but all of them can be easily calculated as described in
previous sections. The results for Regge intercepts for the forward
($t$-channel) and backward ($u$-channel) kinematics are collected in
Table~\ref{table1} for the final $q\bar{q}$ and in
Table~\ref{table2} for the final $l\bar{l}$.

\begin{table}[htbp]
\caption{Rightmost zeros $x_0$ of parabolic cylinder functions $D_p(x)$
determining the values of the leading singularities $\omega_0$ of different
Mellin transform amplitudes $F_r(\omega)$ for $e^+e^-\to q\bar{q}$
annihilation in forward and backward kinematics.}

\begin{tabular}{|c||c|c|c|c||c|c|c|c||}
\hline
$F_r(\omega)$ & \multicolumn{4}{c||} {forward kinematics} & \multicolumn{4}{c||} {backward kinematics} \\
\cline{2-9}
     & $p$    & $x_0$                & $\lambda$ & $\omega_0$
     & $p$    & $x_0$                & $\lambda$ & $\omega_0$                \\
\hline
$F_{LLS}$ & 5.796  & 3.23                   &0.026& 0.083
          & 24.68  & 8.65                   &0.013& 0.111                    \\
$F_{LLT}$ &-0.129  &-2.52$\pm$1.62$\,\imath$&0.106&-0.267$\pm$0.171$\,\imath$
          &-0.805  &-1.98$\pm$2.62$\,\imath$&0.039&-0.076$\pm$0.101$\,\imath$\\
$F_{LRu}$ &-0.083  &-2.65$\pm$1.48$\,\imath$&0.077&-0.205$\pm$0.115$\,\imath$
          & 0.124  &-1.85                   &0.063&-0.117\\
$F_{LRd}$ & 0.062  &-2.25                   &0.063&-0.142
          &-0.05   &-2.81$\pm$1.35$\,\imath$&0.071&-0.199$\pm$0.095$\,\imath$\\
$F_{RL} $ &-0.042  &-2.87$\pm$1.31$\,\imath$&0.077&-0.222$\pm$0.102$\,\imath$
          & 0.05   &-2.365                  &0.071&-0.167\\
$F_{RRu}$ &-0.24   &-2.33$\pm$1.86$\,\imath$&0.064&-0.15 $\pm$0.12 $\,\imath$
          & 6.     & 3.32                   &0.013&0.0428\\
$F_{RRd}$ & 0.75   &-0.34                   &0.026&-0.009
          &-0.188  &-2.40$\pm$1.76$\,\imath$&0.051&-0.124$\pm$0.090$\,\imath$\\
\hline
\end{tabular}
\label{table1}
\end{table}

\begin{table}[htbp]

\caption{Rightmost zeros $x_0$ of parabolic cylinder functions $D_p(x)$
determining the values of the leading singularities $\omega_0$ of different
Mellin transform amplitudes $F_r(\omega)$ for $e^+e^-\to l\bar{l}$
annihilation in forward and backward kinematics. Notations for
isodublet components of $l$ are taken as for muon doublet.}

\begin{tabular}{|c||c|c|c|c||c|c|c|c||}
\hline
$F_r(\omega)$ & \multicolumn{4}{c||} {forward kinematics} & \multicolumn{4}{c||} {backward kinematics} \\
\cline{2-9}
     & $p$    & $x_0$                & $\lambda$ & $\omega_0$
     & $p$    & $x_0$                & $\lambda$ & $\omega_0$                \\
\hline
$F_{LLS}$ &$\infty$& ---                    & --- & 0.132
          & 2.41   & 1.32                   &0.039& 0.051\\
$F_{LLT}$ &-0.090  &-2.63$\pm$1.50$\,\imath$&0.103&-0.270$\pm$0.154$\,\imath$
          &-0.602  &-2.06$\pm$2.40$\,\imath$&0.053&-0.109$\pm$0.127$\,\imath$\\
$F_{LR\nu}$ & ---    & ---                    & --- & ---
          & ---    & ---                    & --- & --- \\
$F_{LR\mu}$ & 0.172  &-1.64                   &0.066&-0.108
          &-0.102  &-2.59$\pm$1.54$\,\imath$&0.086&-0.221$\pm$0.132$\,\imath$\\
$F_{RL} $ & 0.172  &-1.64                   &0.066&-0.108
          &-0.102  &-2.59$\pm$1.54$\,\imath$&0.086&-0.221$\pm$0.132$\,\imath$\\
$F_{RR\nu}$ & ---    & ---                    & --- & ---
          & ---    & ---                    & --- & --- \\
$F_{RR\mu}$ &$\infty$& ---                    & --- & 0.077
          &-0.25   &-2.32$\pm$1.88$\,\imath$&0.077&-0.179$\pm$0.145$\,\imath$\\
\hline
\end{tabular}
\label{table2}
\end{table}

Evidently, in far asymptotics the leading contribution for $W^{\pm}$
production comes from $F_1$ (isotriplet) of the backward $e^+_L e^-_L\to
q_L\bar{q}_L$ whereas the leading contribution to $(Z,\gamma)$
production comes from $F_0$ (isosinglet) of the forward $e^+_L e^-_L\to
q_L\bar{q}_L$. However differences between the non-leading and leading
intercepts are small, and one can expect the role of the first to be
essential at real energies scales.  Moreover, the effects of non-leading
intercepts of the same amplitude can be also large enough at real
energies. Therefore it seems reasonable to numerically compute the
energy dependent amplitudes,  $M_r$, by taking the inverse transform of
the IREE solutions $F_r(\omega)$,  and to calculate  with them the
inelastic cross sections of boson production in central region (in cmf)
with $k_{\perp}^2\sim\mu^2=M_Z^2$.  It seems also suitable to sum over
the final $q\bar{q}$ or  $l\bar{l}$ isotopic states, fixing only the
emitted boson isotopic state.

Easy but cumbersome calculations
lead to the following results.  $W^{\pm}$ production in the
forward and backward kinematics is described by the same formula
(though with different amplitudes involved!):
\begin{equation}
\label{csw}
\frac{\sigma(W^{\pm})}{\sigma_0} = g^2 \left[ |M_{LLT}|^2 + |M_{RL}|^2
\right]~,
\end{equation}
where $\sigma_0$ is the common Born cross section of the elastic process
(see \cite{ggfl}),
$M_{LLT}$ denotes $M_2$ amplitude of $e_L^+e_L^-\to q_L\bar{q}_L$
and $M_{RL}$ denotes the amplitude of $e_R^+e_R^-\to q_L\bar{q}_L$
(and similar for annihilation to leptons).
Let us remind that $W^{\pm}$ are produced first as isovector
boson $A_1$, $A_2$ states and then transform to observable
boson states. In contrast, $(Z, \gamma)$ are being produced first as
isoscalar $B$ or isovector $A_3$ fields, and then transform to the
observable states, $Z$ mainly comes from $A_3$ and $\gamma$ - from $B$.
Easy but cumbersome calculation lead to the following cross
sections for production of $B$ and $A_3$ bosons:
\begin{equation}
\label{cszg}
\sigma(Z,\gamma) = \sigma(A_3) + \sigma(B)~,
\end{equation}

\begin{equation}
\label{csa3}
\frac{\sigma(A_3)}{\sigma_0} = g^2 \left[ |M_{LLT}|^2 +
\frac14\left(|M_{LRu}|^2+|M_{LRd}|^2\right)  + \frac12
|M_{RL}|^2 \right]
\end{equation}
where again the amplitudes $M$ involved are either forward or backward
amplitudes: $M_{LLT} = M_2$ of $e_L^+e_L^-\to q_L\bar{q}_L$,
$M_{LRu}$ stands for $e_L^+e_L^-\to u_R\bar{u}_R$,
$M_{LRd}$ stands for $e_L^+e_L^-\to d_R\bar{d}_R$ and $M_{RL}$ for
$e_R^+e_R^-\to q_L\bar{q}_L$,
and
\begin{eqnarray}
\label{csb}
\frac{\sigma(B)}{\sigma_0} &=& g^{'2} \left[
\frac{(Y_{e_L}\!\mp\!Y_{q_L})^2}{4} \left( \frac14 |M_{LLS}|^2 +
\frac54 |M_{LLT}|^2 + \frac12 |M_{LLS} M_{LLT}| \right)
+ \frac{(Y_{e_L}\!\mp\!Y_{u_R})^2}{4} |M_{LRu}|^2 \right. \nonumber\\
&+& \frac{(Y_{e_L}\!\mp\!Y_{d_R})^2}{4} |M_{LRd}|^2
+ \left. \frac{(Y_{e_R}\!\mp\!Y_{q_L})^2}{4} (2 |M_{RL}|^2)
+ \frac{(Y_{e_R}\!\mp\!Y_{u_R})^2}{4} |M_{RRu}|^2
+ \frac{(Y_{e_R}\!\mp\!Y_{d_R})^2}{4} |M_{RRd}|^2
\right]~,
\end{eqnarray}
where "$\mp$" denotes that "-" sign corresponds to forward amplitudes
and "+" sign to backward amplitudes denoted above.

The same formulae can be used for $e^+e^-\to l\bar{l}$ annihilation 
channel: one has to substitute the appropriate amplitudes $M_r$ and to 
replace electro-weak charge $Y_q$ with the appropriate $Y_l$. 

\begin{figure}[htbp]
\begin{center}
\begin{picture}(240,200)
\put(0,10){
\includegraphics{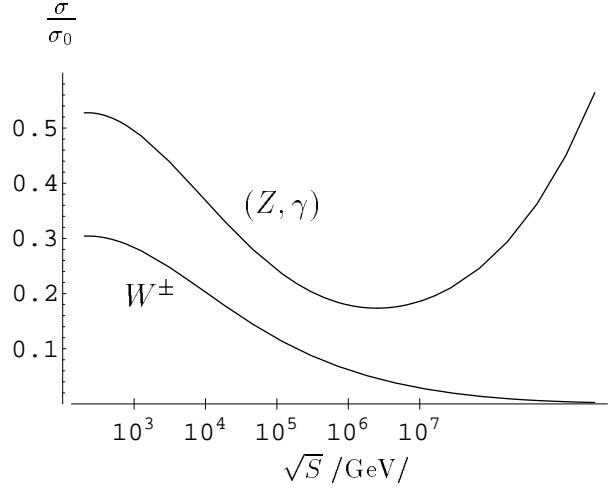}
}
\end{picture}
\end{center}
\caption{Dependence of exclusive $W^{\pm}$ and $(Z,\gamma)$
production on the total energy of $e^+e^-$ annihilation. The
cross sections are divided by the differential elastic Born cross
section $\sigma_0$ to make differences in energy dependencies more
clear.}
\label{fig4}
\end{figure}

\begin{figure}[htbp]
\begin{center}
\begin{picture}(240,200)
\put(0,10){
\includegraphics{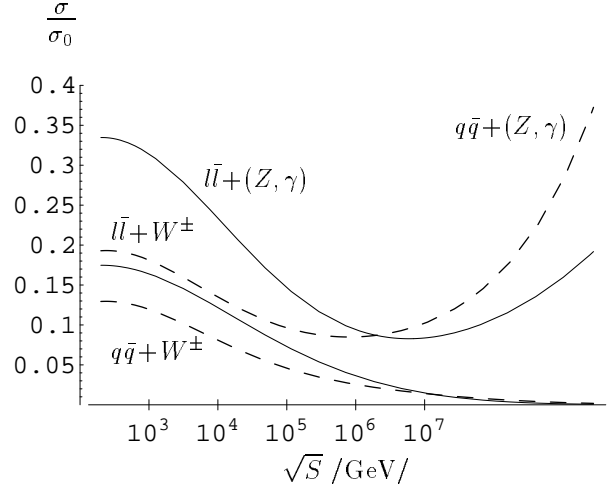}
}
\end{picture}
\end{center}
\caption{Total energy dependence of $W^{\pm}$ and $(Z,\gamma)$
production in different channels of $e^+e^-$ annihilation:
$e^+e^-\to l\bar{l}$ -- solid curves and
$e^+e^-\to q\bar{q}$ -- dashed curves.}
\label{fig5}
\end{figure}

As the Regge kinematics is dominating in the cross sections, we sum 
the contributions of forward and backward kinematics in what follows.
The results of numerical calculations presented  in
Figs.~\ref{fig4},~\ref{fig5} show that
at energies $\sqrt{s} < 10^6$~GeV  $W^{\pm}$ and $(Z,\gamma)$ are
mainly produced in $e^+e^-\to l\bar{l}$ annihilation.  And only at
$\sqrt{s} > 10^6 \div 10^7$~GeV their yields from $e^+e^-\to q\bar{q}$
annihilation become greater (see Fig.~\ref{fig5}).


The explicit asymptotical dominance of exclusive channel
$q\bar{q}+(Z,\gamma)$ over the channel $l\bar{l}+(Z,\gamma)$ stems from
the fact that despite the leading $F_{LLS}$ amplitude in the
Table~\ref{table2} has the greater intercept $\omega_0\approx 0.132$
than $\omega_0\approx 0.111$ of the leading $F_{LLS}$ in the
Table~\ref{table1}, its contribution is multiplied by the zero factor
in Eq.~(\ref{csb}).

\begin{figure}[htbp]
\begin{center}
\begin{picture}(240,200)
\put(0,10){
\includegraphics{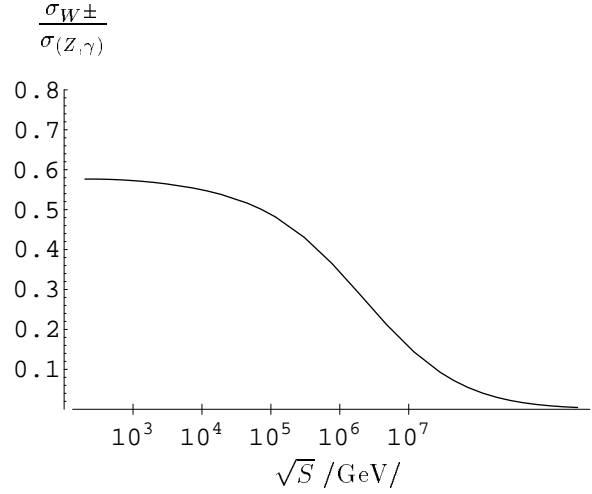}
}
\end{picture}
\end{center}
\caption{Total energy dependence of $W^{\pm}$ to
$(Z,\gamma)$ rate in $e^+e^-$ annihilation.}
\label{fig6}
\end{figure}

The numerical calculation for the ratio of  $W^{\pm}$ to $(Z,\gamma)$
production summed over the both annihilation channels  $e^+e^-\to
l\bar{l}$ and $e^+e^-\to q\bar{q}$ is shown in Fig.~\ref{fig6}.
Thus the DLA predicts rather slow energy dependence of the ratio till
$\sqrt{s}\sim 10^4$GeV and then its relatively rapid decrease.

\begin{figure}[htbp]
\begin{center}
\begin{picture}(240,200)
\put(0,10){
\includegraphics{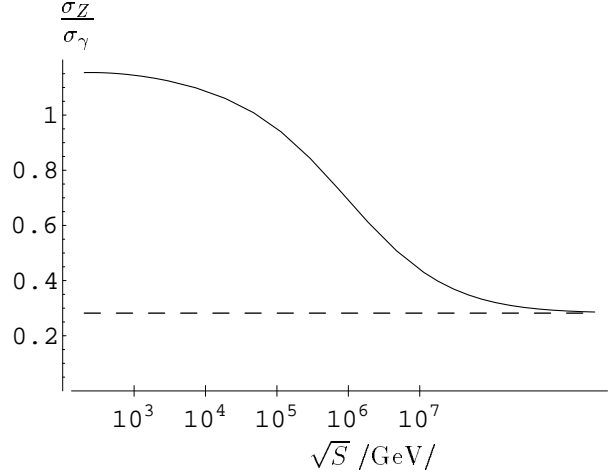}
}
\end{picture}
\end{center}
\caption{Total energy dependence of $Z$ to
$\gamma$ rate in $e^+e^-$ annihilation. The dashed line shows the
asymptotical value of the ratio: $\tan^2\theta_W\approx 0.28$~.}
\label{fig7}
\end{figure}

The energy dependence of the ratio $Z$ to $\gamma$,
\begin{equation}\label{ztog}
\frac{\sigma(Z)}{\sigma(\gamma)} =
\frac{\sigma(A_3) +
\sigma(B)\tan^2\theta_W}{\sigma(A_3)
\tan^2\theta_W + \sigma(B)}~,
\end{equation}
is shown in Fig.~\ref{fig7}. In far asymptotics radiation of
isoscalar field $B$ dominates over radiation of isovector field $A$ and
the ratio tends to the fixed value $\tan^2\theta_W\approx 0.28$.

Worthwhile to note that apart from the results of \cite{ggfl} for
pure QED, the figures of the Table~\ref{table2} show that backward
$e^+_L e^-_L\to l_L\bar{l_L}$ amplitude (i.e. when
antilepton follows the direction of initial electron) for isoscalar
channel in EW theory has the positive intercept though small
enough if compared to forward annihilation amplitude.

Let us emphasise that the demonstrated in Fig.~\ref{fig7} excess of
$Z$ production over $\gamma$ production at $\sqrt{s} < 10^3 \div 10^5$~GeV
as well as excess of $(Z,\gamma)$ emission over $W^{\pm}$ emission in
the same energy range shown in Fig.~\ref{fig4}, and the dominance of
$l\bar{l}$ channel over $q\bar{q}$ channel shown in Fig.~\ref{fig5} may all
happen to be just artifacts of the DLA. To get more reliable predictions for
the cross sections one has at least to account for single logarithmic
corrections as well. The presented figures show that account of
the non-leading DLA effects can make observation of the theoretically
correct predictions (\ref{gammaz}) and (\ref{sigmawt}) hardly possible
even in far future.

\newpage
\section{Conclusion}
\label{CONCLUSION}

In the present paper we have obtained explicit expressions for the 
scattering 
amplitudes for the $e^+e^-$ annihilation into quarks and into leptons   
at the annihilation 
energies $\sqrt{s} \gg 100$~Gev  accompanied by emission of $n$ 
electroweak bosons in the 
multi-Regge kinematics, i.e. in the kinematics 
 where the final particles are in cones with 
opening angles $\ll 1$  around the initial $e^+e^-$ beams. We accounted 
for the double-logarithmic contributions to this process to all orders in the 
EW couplings. We have shown that it is convenient to 
calculate  amplitudes of this process in terms of the 
isoscalar and of the 
isovector amplitudes.      
The isoscalar amplitudes describe production of the 
isoscalar gauge fields. They are controlled by $n + 1$ isoscalar  
Reggeons propagating in the crossing channel. 
The leading intercepts of these Reggeons are 
positive ($\Delta_{S'} = 0.11$ 
and $\Delta_{S} = 0.08$) and therefore such scattering amplitudes grow 
when $s$ increases.         
The isovector amplitudes bring 
sub-leading contributions to the production of the isoscalar bosons and 
in the same time give the leading contributions to production of the 
isovector gauge fields. They are 
governed by $n + 1$ isovector Reggeons with negative 
intercepts $\Delta_{V'} = -0.08$ 
and $\Delta_{V} = -0.27$. It means 
that the amplitudes for isovector production 
decreases when $s$ grows.    
These results lead in particular to the 
fact that production of each $Z$ boson is always accompanied by production 
of a hard photon with the same energy $\gg 100$~Gev. In DLA, 
such hard photons are never produced without $Z$ bosons. The cross sections of 
production of these photons and the $Z$ bosons have  
identical energy dependence, however they are different numerically  
due to difference in the couplings. They are related by 
Eq.~(\ref{gammazn}) at asymptotically high energies ($\geq 10^7$~Gev). The 
$s$ -dependence of the ratio $\sigma^{nZ}/\sigma^{n\gamma}$ for 
lower energies is given in Fig.~\ref{fig7}.      
The energy dependence of 
cross section for the $W$ production is weaker than the one for the 
photons and the $Z$ bosons by factor $s^{-0.36}$ at asymptotically high 
energies. The $s$ -dependence of  these cross sections is shown in 
Figs.~\ref{fig4}-\ref{fig6}. Through this 
paper we consider only the monotonically ordered multi-Regge kinematics 
(\ref{orderti}) and (\ref{orderui}). 
Accounting for the other kinematics can be done in a similar way. 
Though it is likely to bring corrections to explicit  
formulae for the invariant amplitudes $M_r^{(n)}$, 
it cannot change the asymptotic
 relation of Eq.~(\ref{gammazn}) and the fact 
(see Eq.~(\ref{gammawn})) that $\sigma^{(n\gamma)}/\sigma^{(nW)}$ 
decreases with $s$.  

\section{Acknowledgement}

We are grateful to A.~Barroso for interesting discussions. \\ 
The work is supported by grants CERN/FIS/43652/2001, INTAS-97-30494,
SFRH/BD/6455/2001 and RFBR~00-15-96610.

\end{document}